\documentclass[12pt]{article}
\def\blambda{\lambda\!\!\!\raise.55ex\hbox{\,{-}\,}}
\begin{document}
\thispagestyle{empty} \vspace*{2cm}
\begin{center}
\textbf{\Huge Nanoscience\\
 with Attosecond Laser Pulses}

\vspace{1cm}

{\large Janina Marciak-Kozlowska$^{\rm a}$ and  Miroslaw
Kozlowski$^b$}
\end{center}

\vspace{1cm}

$^a${\it Institute of Electron Technology, Al.~Lotnik\'{o}w~32/46
02--668~Warsaw Poland}

$^b${\it Physics Department, Science Teachers College  and
Institute of Experimental Physics,
 Warsaw University
Ho\.{z}a 69, 00-681 Warsaw Poland,\\ e-mail:~mirkoz@fuw.edu.pl}

\newpage
\begin{abstract}
In this paper the interaction of attosecond laser pulses with
matter is investigated. The scattering and potential motion of
heat carriers as well as the external force are considered.
Depending on the ratio of the scatterings and potential motion the
heat transport is described by the thermal forced Klein-Gordon or
thermal modified telegraph equation. For thermal Heisenberg type
relation $V\tau\sim\hbar$ ($\tau$ is the relaxation time and $V$
is the potential) the heat transport is described by the thermal
distortionless damped wave equation. In this paper Klein-Gordon,
modified telegraph\footnote{Considering history of the formulation
and application of the hyperbolic partial differential equation,
Olivier Heaviside for the first time write (in 1876 ) hyperbolic
equation for voltage $v(x,t)$ in submarine cable. [G.~Thomas and
D.~J.~Raine, \textit{Physics to a degree}, Gordon and Breach,
2000, p.~43]. $\frac{\partial^2v(x, t)}{\partial
x^2}=RC\frac{\partial v(x,t)}{\partial t}+LC\frac{\partial
^2v}{\partial t^2}$. We propose the name \textit{Heaviside
equation} instead of the \textit{telegraph equation} for the
hyperbolic partial equation (thermal wave equation with both first
and second derivative).} equation and wave equation are solved.

Key words: Attosecond laser pulses; Quantum heat transport
equation; Klein-Gordon thermal equation.
\end{abstract}

\newpage
\section{Introduction}
        With attosecond lasers (1~as $=10^{-18}$~s) physicist and engineers
        are
        now close to controlling the motion of electrons on a timescale
that is
        substantially shorter than the oscillation period of visible light.
It is also possible with attosecond laser pulse to rip an electron
wave-packet from the core of an atom and set it free with similar
temporal precision.

        Since we know that there is an absolute velocity, velocity of light
it is sufficient to define the smallest time interval (e.g. 1~as) and the
unit of length is a derivative quantity ($3\cdot 10^{-10}~{\rm
m}=0.3~{\rm nm} =$ Bohr radius).

        After the standards of time and space are defined the laws of classical
physics relating such parameters as distance, time, velocity, temperature
are assumed to be independent of accuracy with which these parameters can
be measured. It should be noted that this assumption does not enter
explicitly into the formulation of the classical physics. This means that
together with the assumption of the existence of the objective reality
independent of any measurements (in classical physics) it was tacitly
assumed that \textit{there is a possibility of unlimited increase of
measurements accuracy}. Bearing in mind the atomicity of time i.e.
considering the smallest time period, Planck time the above statement is
obviously not true. With attosecond laser pulses we are at the limit of
laser time resolution.

        With attosecond laser pulses we enter new Nano World where size become
comparable to atomic dimension, where transport phenomena follows
different laws than in macro world. This means the end of first step
miniaturization (from $10^{-3}$ m to $10^{-6}$ m) and the beginning new
one (from $10^{-6}$ to $10^{-9}$). Nano World is a quantum world with all
predicable and nonpredicable (yet) features.

        In this paper we develop and solve the quantum relativistic heat
transport equation for Nano World transport phenomena when external
forces exist. This is the generalization of the results of paper~\cite{1}
in which quantum relativistic hyperbolic equation was proposed and
solved.

        \section{Relativistic hyperbolic heat transport equation}
        In paper~\cite{1} relativistic hyperbolic transport equation (RHT)
was formulated:
        \begin{equation}
        \frac{1}{v^2}\frac{\partial^2 T}{\partial
    t^2}+\frac{m_0\gamma}{\hbar}\frac{\partial T}{\partial t}
    =\nabla^2T.\label{eq1}
        \end{equation}
In equation~(\ref{eq1}) $v$ is the velocity of heat waves, $m_0$
is the mass of heat carrier and $\gamma$ -- the Lorentz factor
$\gamma=(1-\frac{v^2}{c^2})^{-1/2}$. As was shown in
paper~\cite{1} the heat energy (\textit{heaton temperature}) $T_h$
can be defined as follows:
        \begin{equation}
        T_h=m_0\gamma v^2.\label{eq2}
        \end{equation}
Considering that $v$, thermal wave velocity equals~\cite{1}
        \begin{equation}
        v=\alpha c,\label{eq3}
        \end{equation}
where $\alpha$ is the coupling constant for the interactions which
generate the \textit{thermal wave} ($\alpha=1/137$ and
$\alpha=0.15$ for electromagnetic and strong force respectively)
\textit{heaton temperature} equals
        \begin{equation}
        T_h=\frac{m_0\alpha^2c^2}{\sqrt{1-\alpha^2}}.\label{eq4}
        \end{equation}
        From formula~(\ref{eq4}) one concludes that \textit{heaton temperature} is the
linear function of the mass $m_0$ of the heat carrier. It is quite
interesting to observe that the proportionality of $T_h$ and heat carrier
mass $m_0$ was the first time observed in ultrahigh energy heavy ion
reactions measured at CERN~\cite{2}. In paper~\cite{2} it was shown that
temperature of pions, kaons and protons produced in Pb+Pb, S+S reactions
are proportional to the mass of particles. Recently at Rutherford
Appleton Laboratory (RAL) the VULCAN laser was used to produce the
elementary particles: electrons and pions~\cite{3}.

        In the present paper the forced relativistic heat transport equation
will be studied and solved. In paper~\cite{4} the damped thermal wave
equation was developed:
        \begin{equation}
        \frac{1}{v^2}\frac{\partial^2 T}{\partial
t^2}+\frac{m}{\hbar}\frac{\partial T}{\partial
t}+\frac{2Vm}{\hbar^2}T-\nabla^2T=0.\label{eq5}
        \end{equation}
        The relativistic generalization of equation~(\ref{eq5}) is quite
obvious:
        \begin{equation}
        \frac{1}{v^2}\frac{\partial^2 T}{\partial
        t^2}+\frac{m_0\gamma}{\hbar}\frac{\partial T}{\partial t}+
        \frac{2Vm_0\gamma}{\hbar^2}T-\nabla^2T=0.\label{eq6}
        \end{equation}
        It is worthwhile to note that in order to obtain nonrelativistic
equation we put $\gamma=1$.

If the nucleus contains $Z$ protons than we have for electron temperature
in atom
        \begin{equation}
        T^Z_h({\rm atom})=m_e(Z\alpha)^2c^2.\label{e22}
        \end{equation}
        Equation~(\ref{e22}) also tell us the magnitude of the temperature of
the ionization of the atom:
        $$
        T^Z_h({\rm
ionization})=\frac12m_ec^2\left(Z\alpha\right)^2
\left(1+\frac{m_e}{m_p}\right)^{-1}\sim Z^210^5{\rm K}.
        $$
        The motion of electron in the atom is equivalent to the flow of an
electric current in a loop of wire. With attosecond laser pulses
we will be able to influence the current in the atomic ``wire''.
This opens quite new perspective for the Nano World technology or
\textit{nanotechnology}. The new equation~(\ref{eq6}) is the
natural candidate for the master equation which can be used in
\textit{nanotechnology}.

When external force is present $F(x,t)$ the forced damped heat transport
is obtained instead of equation~(\ref{eq6}) (in one dimensional case):
        \begin{equation}
        \frac{1}{v^2}\frac{\partial^2 T}{\partial
t^2}+\frac{m_0\gamma}{\hbar}\frac{\partial T}{\partial
t}+\frac{2Vm_0\gamma}{\hbar^2}T-\frac{\partial^2T}{\partial
x^2}=F(x,t).\label{eq7}
        \end{equation}
        The hyperbolic relativistic quantum heat transport equation (RQHT),
formula~(\ref{eq7}), describes the forced motion of heat carriers which
undergo the scatterings $(\frac{m_0\gamma}{\hbar}\frac{\partial
T}{\partial t}\;{\rm term})$ and are influenced by potential
$(\frac{2Vm_0\gamma}{\hbar^2}T\;{\rm term})$.

        The solution of equation can be written as
        \begin{equation}
        T(x,t)=e^{-t/2\tau}u(x,t),\label{eq8}
        \end{equation}
        where $\tau=\hbar/(mv^2)$ is the relaxation time. After substituting
formula~(\ref{eq8}) to the equation~(\ref{eq7}) we obtain new equation
        \begin{equation}
        \frac{1}{v^2}\frac{\partial^2 u}{\partial t^2}-\frac{\partial^2
u}{\partial x^2}+qu(x,t)=e^{\frac{t}{2\tau}}F(x,t),\label{eq9}
        \end{equation}
        and
        \begin{eqnarray}
        q&=&\frac{2Vm}{\hbar^2}-\left(\frac{mv}{2\hbar}\right)^2,\label{eq10}\\
        m&=&m_0\gamma.\nonumber
        \end{eqnarray}
        Equation~(\ref{eq9}) can be written as:
        \begin{equation}
        \frac{\partial^2 u}{\partial t^2}-v^2\frac{\partial^2 u}{\partial x^2}+
        qv^2u(x,t)=G(x,t),\label{eq11}
        \end{equation}
        where
        $$
        G(x,t)=v^2e^{\frac{t}{2\tau}}F(x,t).
        $$
        When $q>0$ equation~(\ref{eq11}) is the forced Klein-Gordon (K-G)
equation. The solution of the forced Klein-Gordon equation for the
initial conditions:
        \begin{equation}
        u(x,0)=f(x) \qquad u_t(x,0)=g(x)\label{eq12}
        \end{equation}
        has the form:~\cite{4}
        \begin{eqnarray}
        u(x,t)&=&\frac{f(x-vt)+f(x+vt)}{2}\label{eq14} \\
        &&\mbox{}+\frac{1}{2v}\int\limits^{x+vt}_{x-vt}{g(\zeta)J_0
        \left[q\sqrt{v^2t^2-(x-\zeta)^2}\right]d\zeta}\nonumber \\
        &&\mbox{}-\frac{\sqrt{q}\; vt}{2}\int\limits^{x+vt}_{x-vt}{f(\zeta)
        \frac{J_1\left[q\sqrt{v^2t^2-(x-\zeta)^2}\right]}
        {\sqrt{v^2t^2-(x-\zeta)^2}}d\zeta}\nonumber \\
        &&\mbox{}+\frac{1}{2v}\int\limits^t_0\int\limits^{x+v(t-t')}_{x-v(t-t')}
        {G(\zeta,t')J_0\left[q\sqrt{v^2(t-t')^2-(x-\zeta)^2}\right]dt'd\zeta}.\nonumber
        \end{eqnarray}
        When $q<0$ equation~(\ref{eq11}) is the forced modified Heaviside
(telegraph) equation with the solution:~\cite{4}
        \begin{eqnarray}
        u(x,t)&=&\frac{f(x-vt)+f(x+vt)}{2}\label{eq15} \\
        &&\mbox{}+\frac{1}{2v}\int\limits^{x+vt}_{x-vt}{g(\zeta)I_0
        \left[-q\sqrt{v^2t^2-(x-\zeta)^2}\right]d\zeta}\nonumber \\
        &&\mbox{}+\frac{v\sqrt{-q}\;t}{2}\int\limits^{x+vt}_{x-vt}{f(\zeta)
\frac{I_1\left[-q\sqrt{v^2t^2-(x-\zeta)^2}\right]}
{\sqrt{v^2t^2-(x-\zeta)^2}}}d\zeta\nonumber \\
        &&\mbox{}+\frac{1}{2v}\int\limits^{t'}_0\int\limits^{x+v(t-t')}_{x-v(t-t')}
{G(\zeta,t')I_0\left[-q\sqrt{v^2(t-t')^2-(x-\zeta)^2}\right]d\zeta
dt'}.\nonumber
        \end{eqnarray}
        When $q=0$ equation~(\ref{eq11}) is the forced thermal
equation~\cite{4}.
        \begin{equation}
        \frac{\partial^2 u}{\partial t^2}-v^2\frac{\partial^2 u}{\partial
x^2}=G(x,t).\label{eq16}
        \end{equation}
        On the other hand one can say that equation~(\ref{eq16}) is the
distortionless hyperbolic equation. The condition $q=0$ can be rewrite
as:
        \begin{equation}
        V\tau=\frac{\hbar}{8}.\label{eq17}
        \end{equation}
        The equation~(\ref{eq17}) is the analogous to the Heisenberg
uncertainty relations. Considering formula~(\ref{eq2})
equation~(\ref{eq17}) can be written as:
        \begin{equation}
        V=\frac{T_h}{8}, \qquad V<T_h.\label{eq18}
        \end{equation}
        One can say that the distortionless waves can be generated only if
$T_h>V$. For $T_h<V$, i.e. when the ``Heisenberg rule'' is broken, the
shape of the thermal waves is changed.

        We consider the initial and boundary value problem for the
inhomogenous thermal wave equation in semi-infinite interval~\cite{4}:
that is
        \begin{equation}
        \frac{\partial^2 u}{\partial t^2}-v^2\frac{\partial^2 u}{\partial x^2}=
        G(x,t), \qquad 0<x<\infty, \qquad t>0,\label{eq19}
        \end{equation}
        with initial condition:
        $$
        u(x,0)=f(x), \qquad \frac{\partial u(x,0)}{\partial t}=g(x),
\qquad 0<x<\infty,
        $$
        and boundary condition
        \begin{equation}
        au(0,t)-b\frac{\partial u(0,t)}{\partial x}=B(t),\qquad t>0,\label{eq20}
        \end{equation}
        where $a\geq0,\,b\geq0, \,a+b>0$ (with $a$ and $b$ both equal to
constants) and $F,\, f, \, g$ and $B$ are given functions. The solution
of equation~(\ref{eq19}) is of the form~\cite{4}:
        \begin{eqnarray}
        u(x,t)&=&\frac12\left[f(x-vt)+f(x+vt)\right]+\frac{1}{2v}
        \int\limits^{x+vt}_{x-vt}{g(s)ds}\\ \nonumber
        &&\mbox{}+\frac{1}{2v}\int\limits^{t}_0\int\limits^{x+v(t-t')}_{x-v(t-t')}
        {F(s,t')ds dt'}.\label{eq21}
        \end{eqnarray}
    In the special case where $f=g=F=0$ we obtain the following solution
of the initial and boundary value problem~(\ref{eq19},~\ref{eq20}):
    \begin{eqnarray}
  u(x,t)=\left\{\begin{array}{lcl}
    0, &\quad & x>vt,\\
        {\displaystyle\frac{v}{b}\int\limits_0^{t-\frac{x}{v}}{\exp\left[\frac{va}{b}
\left(y-t+\frac{x}{v}\right)
    \right]B(t)dt},} & \quad &0<x<vt,\end{array}\label{eq99} \right.
    \end{eqnarray}
if $b\neq0$. If $a=0$ and $b=1$, we have:
    \begin{eqnarray}
    u(x,t)=\left\{
    \begin{array}{lcl}0, & \quad & x>vt,\nonumber\\
    {\displaystyle v\int\limits_0^{t-\frac{x}{v}}B(y)dy}, &\quad& 0<x<vt,\nonumber
    \end{array}\right.
    \end{eqnarray}
whereas if $b=0$ and $a=1$, we obtain:
    \begin{eqnarray}
    u(x,t)=\left\{
    \begin{array}{lcl}0, &\quad& x>vt,\nonumber\\
    {\displaystyle B\left(t-\frac{x}{v}\right)}, &\quad& 0<x<vt.\nonumber
    \end{array}\right.
    \end{eqnarray}

        It can be concluded that the boundary condition~(\ref{eq20}) gives rise
to a wave of the form $K(t-\frac{x}{v})$ that travels to the right with
speed $v$. For this reason the foregoing problem is often referred to as
a \textit{signaling problem} for the thermal waves.

\section{Looking into Nano World}
With attosecond laser pulses we expect to be able to track the motion of
an electron within a femtosecond. We could also follow the relaxation of
the remaining bound electrons to their new equilibrium states. Once we
can generate attosecond pulses with much higher energies we will be able
to use them as both pumps and probes in pump-probe spectroscopy on the
atomic level.

Up to now the pump-probe method allows the investigation of the heat
transport on the bulk scale. The experiments triggered by Brorson
paper~\cite{6} disclosed the finite velocity of the heat transport i.e.
the hyperbolicity of heat transport equation.

Accordingly with attosecond pump-probe method we will be able the study
of heat transport in the Nano World. As was shown in~\cite{1} one can
define the electron temperature (heaton temperature) in ground state
hydrogen like atom:
        \begin{equation}
        T_h({\rm atom}) =m_e\alpha^2c^2,\label{eq22}
        \end{equation}
        and in hydrogen like molecule~\cite{7,8}
        \begin{equation}
        T_h({\rm molecule})=\alpha^2\frac{m_e}{m_p}m_ec^2.\label{eq23}
        \end{equation}
        \section{Conclusion}
        In this paper damped forced thermal wave equation was proposed and
solved. For Cauchy boundary initial condition the solution of the
equation was presented. It was shown that the new equation, depending on
the relation between scattering and potential motion, is of the form of
the forced thermal Klein-Gordon equation or the forced modified Heaviside
(telegraph) equation. For the thermal Heisenberg type relation
$V\tau\sim\hbar/8$ the new equation is the distortionless forced wave
equation. In the context of the forced wave equation the \textit{
signaling\/} solution was obtained.

\end{document}